# Hierarchical Multi-Agent DRL-Based Dynamic Cluster Reconfiguration for UAV Mobility Management

Irshad A. Meer, *Member, IEEE*, Karl-Ludwig Besser, *Member, IEEE*, Mustafa Ozger, *Member, IEEE*, Dominic A. Schupke, *Senior Member, IEEE*, H. Vincent Poor, *Life Fellow, IEEE*, and Cicek Cavdar, *Member, IEEE*

***Abstract*— Multi-connectivity involves dynamic cluster formation among distributed access points (APs) and coordinated resource allocation from these APs, highlighting the need for efficient mobility management strategies for users with multi-connectivity. In this paper, we propose a novel mobility management scheme for unmanned aerial vehicles (UAVs) that uses dynamic cluster reconfiguration with energy-efficient power allocation in a wireless interference network. Our objective encompasses meeting stringent reliability demands, minimizing joint power consumption, and reducing the frequency of cluster reconfiguration. To achieve these objectives, we propose a hierarchical multi-agent deep reinforcement learning (H-MADRL) framework, specifically tailored for dynamic clustering and power allocation. The edge cloud connected with a set of APs through low latency optical back-haul links hosts the high-level agent responsible for the optimal clustering policy, while low-level agents reside in the APs and are responsible for the power allocation policy. To further improve the learning efficiency, we propose a novel action-observation transition-driven learning algorithm that allows the low-level agents to use the action space from the high-level agent as part of the local observation space. This allows the lower-level agents to share partial information about the clustering policy and allocate the power more efficiently. The simulation results demonstrate that our proposed distributed algorithm achieves comparable performance to the centralized algorithm. Additionally, it offers better scalability, as the decision time for clustering and power allocation increases by only 10% when doubling the number of APs, compared to a 90% increase observed with the centralized approach.**

***Index Terms*— Reinforcement learning, energy-efficiency maximization, ultra-reliable communications, UAV communications.**

Received 18 November 2024; revised 24 April 2025 and 14 October 2025; accepted 9 December 2025. Date of publication 19 December 2025; date of current version 15 January 2026. This work was supported in part by the CELTIC-NEXT Project, 6G for Connected Sky (6G-SKY), 3D NETworks for 6G Mobile Communications Applications (3D-NET), and Robust and AI Native 6G for Green Networks (RAI-6Green), with funding received from Vinnova, Swedish Innovation Agency. The work of Karl-Ludwig Besser is supported by Security Link and was supported by the German Research Foundation (DFG) under grant BE 8098/1-1. The work of H. Vincent Poor was supported in part by an Innovation Grant from Princeton NextG. An earlier version of this paper was presented in part at the 2024 IEEE International Conference on Machine Learning for Communication and Networking (ICMLCN) [DOI: 10.1109/ICMLCN59089.2024.10625071]. The associate editor coordinating the review of this article and approving it for publication was W. Xu. *(Corresponding author: Irshad A. Meer.)*

Irshad A. Meer and Cicek Cavdar are with the School of Electrical Engineering and Computer Science, KTH Royal Institute of Technology, 100 44 Stockholm, Sweden (e-mail: iameer@kth.se; cavdar@kth.se).

Karl-Ludwig Besser was with the Department of Electrical and Computer Engineering, Princeton University, Princeton, NJ 08544 USA. He is now with the Department of Electrical Engineering, Linköping University, 581 83 Linköping, Sweden (e-mail: karl-ludwig.besser@liu.se).

Mustafa Ozger is with the School of Electrical Engineering and Computer Science, KTH Royal Institute of Technology, 100 44 Stockholm, Sweden, and also with the Department of Electronic Systems, Aalborg University, 2450 Copenhagen, Denmark (e-mail: mozger@es.aau.dk).

Dominic A. Schupke is with the Central Research and Technology Airbus, 82024 Taufkirchen, Germany (e-mail: dominic.schupke@airbus.com).

H. Vincent Poor is with the Department of Electrical and Computer Engineering, Princeton University, Princeton, NJ 08544 USA (e-mail: poor@princeton.edu).

## I. Introduction

In TRADITIONAL cell-centric wireless networks, each user is associated with a single access point (AP), and mobility management is primarily handled through handovers between cells. However, as networks evolve to support high-mobility users such as unmanned aerial vehicles (UAVs), this model faces significant limitations. Frequent handovers at cell edges often lead to service disruptions and increased signaling overhead. In contrast, user-centric, also called cell-less, architectures decouple mobility from rigid cell boundaries [1], [2]. Instead of connecting to a single AP, each user is dynamically served by a cluster of geographically distributed APs using the same time-frequency resources [3]. Such cooperative transmission can be realized through techniques like coordinated multi-point (CoMP) [4], cloud-radio access networks (C-RANs) [5], and cell-free networks [6]. This multi-connectivity paradigm enables seamless connectivity and robust communication, particularly for UAVs with highly dynamic 3D mobility patterns [2], [4]. However, it also introduces new challenges in mobility management. First, the network must determine the optimal serving cluster of APs for each user to satisfy stringent quality of service (QoS) requirements, such as reliability, under dynamic conditions. In UAV scenarios, this challenge is further amplified by high-mobility in three dimensions, and altitude dependent channel conditions to multiple, and often distant, APs, which complicate clustering decisions. Second, while involving multiple APs enhances reliability, it also increases power consumption due to concurrent transmissions. Therefore, minimizing total transmission power without compromising communication quality becomes a key design objective. Third, as users move, their serving clusters must be updated accordingly. However, frequent or unnecessary reconfigurations lead to increased control overhead and latency. Hence, the network should minimize reconfigurations while ensuring that user QoS remains unaffected.

In addition, developing an efficient power allocation scheme for dynamic clusters in a multi-connectivity wireless interference network presents a significant challenge. This requires continuously adapting to changing network conditions and cluster configurations while managing interference, all under stringent QoS constraints [7]. Different approaches, including optimization theory [8], matching theory [9], [10], and game theory [11] have been explored to address the challenges of dynamic clustering and resource allocation in different networks. However, these conventional techniques are often hindered by several issues. For example, they depend on having complete and real-time information about network dynamics, which is unrealistic in a wireless scenario where channel conditions fluctuate rapidly, especially for UAV communication having a probabilistic line of sight conditions with ground APs. Additionally, These methods are computationally intensive and struggle to scale, with complexity increasing exponentially as network size grows [3].

Machine learning (ML), especially deep reinforcement learning (DRL), has been recognized as a more adaptable and resilient method for managing cluster reconfiguration and resource allocation, by interacting with an unpredictable wireless environment [1], [2], [12], [13]. Through environmental learning, DRL leverages unique characteristics of communication networks to learn the desired policies. While a centralized DRL approach can efficiently solve the cluster reconfiguration and resource allocation problem [1], it faces scalability issues as the network size increases. On the other hand, multi-agent deep reinforment learning (MADRL) addresses scalability by enabling distributed decision-making. However, despite its scalability advantages, MADRL may face performance limitations when agents have only partial observations. Although value decomposition methods [14], [15] enable decentralized cooperation, they are typically designed for fully cooperative tasks with shared goals. In contrast, our problem involves heterogeneous and tightly coupled objectives, such as optimal clustering and power control, which are better addressed through a hierarchical framework, where clustering is coordinated at the edge cloud and power control is distributed among AP agents. As a result, both approaches present distinct advantages and challenges that must be balanced. This motivates the use of hierarchical MADRL, which combines local decision-making with a higher-level coordination structure, offering a potential solution to both scalability and performance trade-offs.

In this paper, we provide a scalable mobility management framework for the multi-connectivity scenario, which meets stringent reliability requirements while minimizing transmit power and the number of cluster reconfigurations, even when the network size increases. Given the uncertainties in UAVs mobility and the variability of channel conditions over time, we model the joint problem of cluster reconfiguration and power allocation as a Markov decision process (MDP) and propose a hierarchical MADRL framework to solve it. A high-level agent operates within the edge cloud and makes the clustering decision, while multiple low-level agents each associated with individual AP perform power allocation to the assigned users, as shown in Figure 1. The core concept involves decentralizing the decision-making process, assigning responsibilities to different levels of the network rather than relying on a single decision-making agent. To improve the performance of hierarchical MADRL and provide the low level agents with global environmental information for better decision making, we propose a novel action-observation transition mechanism. This mechanism allows the action from the higher agent to be used as a part of the observation space of the low level agents.

The main contributions are summarized below.

- We formulate an optimization problem focused on dynamic clustering of APs and their power allocation for UAVs within a wireless interference network. The objective is to satisfy the stringent reliability requirements, specifically considering error probabilities in finite blocklength regimes, reduce power consumption, and minimize the frequency of cluster reconfigurations.
- We propose a hierarchical MADRL approach to enhance communication reliability, lower power consumption, and minimize cluster reconfiguration frequency in a dynamic environment while improving the scalability and reducing complexity and overhead. The high-level agent in the edge cloud, connected to the APs, optimizes the AP clustering strategy, while the low-level AP agents optimize the power allocation. Additionally, the APs' power allocation strategies influence the edge cloud's clustering strategy, further improving system performance.
- We further propose an action-observation transition-driven hierarchical framework, where the observation space of the low-level agents includes the actions of the higher-level agent. This integration ensures that the decisions made by the higher-level agent directly influence and guide the behavior of the low-level agents, creating a more cohesive and adaptive system for multi-connectivity mobility management.
- Finally, we validate the feasibility of our proposed hierarchical MADRL through numerical simulations. The results show that our proposed algorithm can achieve better performance in comparison with other existing works, including the central clustering approach in [1], opportunistic clustering algorithm in [16], and the optimization based joint clustering and power control algorithm from [8].

## II. Related Work

Mobility management for aerial users has been extensively studied in single-connectivity networks. Several works [17], [18], [19], [20], [21] formulate handover control as an optimization task and apply reinforcement learning (RL) to reduce handovers, manage interference, and enhance throughput. However, they focus on single-AP association, without addressing clustering or multi-connectivity.

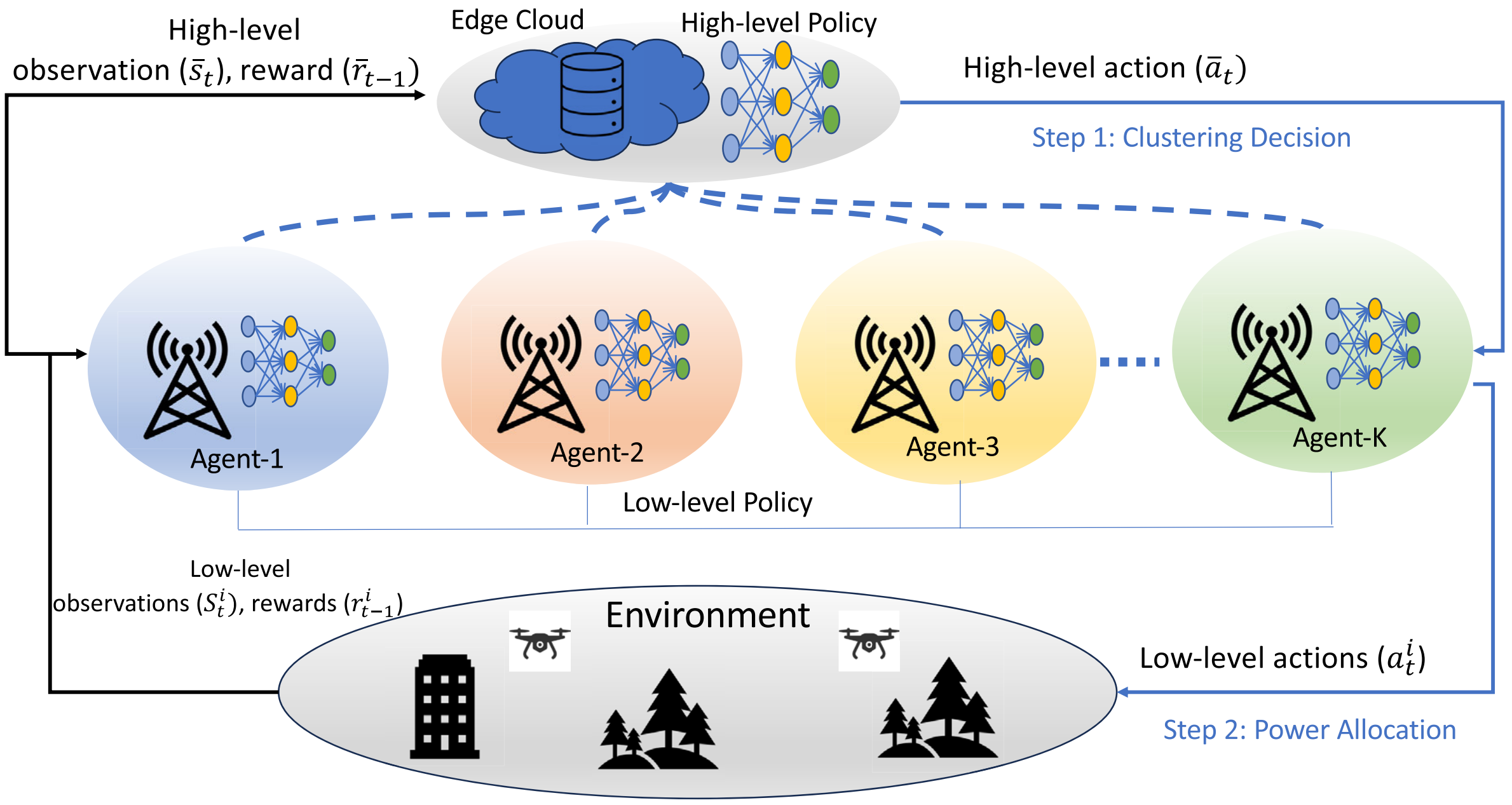

Fig. 1. Network architecture of the proposed H-MADRL. The system consists of $K$ multi-agents at the APs, connected to a high-level agent in the edge cloud. These agents collaboratively serve $N$ aerial users (AUs). The H-MADRL framework operates in two steps: (1) the high-level agent makes a clustering decision and communicates it to the low-level agents; (2) the low-level agents allocate power to the assigned AUs based on the clustering information and local observations.

In multi-connectivity settings, where users connect to multiple APs, clustering and coordination become more complex. The approach in [16] selects clusters based on channel conditions in an open-radio access network (O-RAN) setup, but frequent reconfiguration incurs high signaling overhead. In [22], clustering and beamforming are optimized via RL for terrestrial users, but mobility and reliability are ignored. The work in [23] considers reliability for aerial users, but in a single-user, interference-free scenario. Optimization-based clustering and power control are also explored in [8] where convex approximation is employed to improve system efficiency. However, the proposed approach in [8] is computationally intensive for dynamic environments.

Matching theory has been applied to clustering in static multi-connectivity scenarios [9], [10], but without accounting for user mobility. Game theoretic and learning-based methods have also been employed for multi-connectivity clustering in the existing literature. For instance, in [11], AP selection in cell-free MIMO is modeled as a local altruistic game with convergence guarantees. A double-layer MARL framework is proposed in [12] for energy-efficient clustering in CF XL-MIMO, while [13] presents decentralized and federated MARL for scalable clustering under mobility. To improve scalability and decision efficiency, hierarchical reinforcement learning has been applied in various wireless contexts. In [24], Hierarchical deep reinforcement learning (HDRL) is used for drone trajectory optimization and resource allocation. A similar framework is applied in [25] for joint radio access technologies (RATs) assignment and power control, and in [26] for terrestrial handover and power management. These studies, however, do not address mobility management under reliability constraints in the finite block-length regime.

As summarized in Table I, most existing works address isolated aspects of mobility management in dynamic multi-connectivity scenarios. To the best of our knowledge, there is limited research focusing on energy-efficient mobility management solutions that meet stringent reliability requirements in the finite block-length regime for aerial users with multi-connectivity configurations. This paper fills this gap by providing a unified, hierarchical MADRL framework that jointly addresses clustering and power control for aerial users under stringent reliability requirements.

## III. System Model and Problem Formulation

We consider a downlink communication scenario under an O-RAN architecture, where each UAV is served by a dynamically formed cluster of APs. In this setup, the APs act as transmitters and the UAVs (also referred to as aerial users, AUs) as receivers. A total of $K$ O-RAN radio units (O-RUs) (hereafter referred to as APs) are deployed at fixed ground locations and connected to the edge cloud with virtualization and processing capabilities [27]. Meanwhile, $N$ AUs move through the service area based on a 3D stochastic mobility model from [28]. All the AUs are equipped with a single omnidirectional antenna while each APs is equipped with $L$ uniform planar array antennas. The notation used in the paper are given in the Table II. When AU $i$ enters the coverage area of the edge cloud, a cluster of APs under the clustering strategy $\Gamma$ form the cluster to serve AU $i$. A general clustering strategy is defined as follows.

*Definition 1:* A clustering strategy $\Gamma$ defines a collection $\{\mathcal{M}_1^\Gamma, \mathcal{M}_2^\Gamma, \dots, \mathcal{M}_N^\Gamma\}$ of subsets of $\mathcal{K}$, where $\mathcal{M}i^\Gamma \subseteq \mathcal{K}$ is referred to as the serving cluster for AU $i$, with $|\mathcal{M}_i^\Gamma| \geq 1$ and $|\bigcup_{i=1}^N \mathcal{M}_i^\Gamma| \leq K$.

TABLE I
COMPARISON OF RELEVANT RESEARCH WITH THIS PAPER

| Ref. | H-MARL | Multi Connectivity | Clustering | Power Control | Scalability | Mobility | Reliability |
|---|---|---|---|---|---|---|---|
| **Conf.** [1] | ✘ | ✔ | ✔ | ✔ | ✘ | ✔ | ✔ |
| [2] | ✘ | ✔ | ✔ | ✔ | ✘ | ✔ | ✘ |
| [8] | ✘ | ✔ | ✔ | ✔ | ✘ | ✘ | ✘ |
| [4], [9]–[13] | ✘ | ✔ | ✔ | ✘ | ✔ | ✘ | ✘ |
| [17]–[21] | ✘ | ✘ | ✘ | ✘ | ✔ | ✔ | ✘ |
| [16] | ✘ | ✔ | ✔ | ✘ | ✘ | ✔ | ✘ |
| [22] | ✘ | ✔ | ✔ | ✔ | ✘ | ✘ | ✘ |
| [23] | ✘ | ✔ | ✔ | ✔ | ✘ | ✔ | ✔ |
| [24]–[26] | ✔ | ✘ | ✘ | ✔ | ✔ | ✔ | ✘ |
| **This Paper** | ✔ | ✔ | ✔ | ✔ | ✔ | ✔ | ✔ |

TABLE II
TABLE OF NOTATION

| Notation | Definition |
|---|---|
| $\mathcal{K}, k, K$ | The set, the index and the number of APs |
| $\mathcal{N}, i, N$ | The set, the index and the number of AUs |
| $h_{ik}(t)$ | Channel gain from AP $k$ to AU $i$ at time $t$ |
| $\Gamma$ | Clustering strategy at the edge cloud |
| $\mathcal{N}_k$ | Set of assigned users to AP $k$, $\mathcal{N}_k \subseteq \mathcal{N}$ |
| $\mathcal{M}_i$ | Serving cluster for AU $i$ |
| $P_{T,ik}(t)$ | Transmitted power from AP $k$ to AU $i$ |
| $P_i(t)$ | Received power at user $i$ |
| $P_{max}$ | Maximum allowed transmit power |
| $N_0$ | Noise spectral density |
| $G_0$ | Antenna array gain |
| $\theta_{i,k}(t)$ | Elevation angle between AP $k$ and AU $i$ |
| $\phi_{i,k}(t)$ | Azimuth angle between AP $k$ and AU $i$ |
| $\gamma_i^{\mathcal{M}_i}$ | Receive SINR at AU $i$ served by cluster $\mathcal{M}_i$ |
| $b_i$ | Number of transmitted information bits to user $i$ |
| $n$ | Finite block-length |
| $\varepsilon_i$ | Decoding error probability at AU $i$ |
| $\varepsilon_{\max}$ | Maximum acceptable error probability |
| $\gamma^{th}$ | SINR threshold |
| $O_i(t)$ | SINR outage probability of AU $i$ |
| $H$ | Size of hidden layers in policy/value networks |
| $L$ | Number of layers in the neural networks |

Note that the clusters $\mathcal{M}_i^{\Gamma}$ may overlap, and their union may not necessarily cover all elements of $\mathcal{K}$. Therefore, the total received power $P_i$ at AU $i$ at time $t$ is given as

$$P_i(t) = \sum_{k=1}^{|\mathcal{M}_i^{\Gamma}(t)|} h_{ik}(t)\, P_{T,ik}(t)\, G\big(\theta_{i,k}(t), \phi_{i,k}(t)\big)\,, \tag{1}$$

where $P_{T,ik}$ denotes the transmit power of AP $k$ to user $i$, and $h_{ik}$ is the power attenuation between APs $k$ and user $i$, i.e., the combined path loss and fading effects. These effects are modeled according to the 3GPP specifications for AUs [29], which account for height-dependent channel conditions. With known location of the user, we incorporate the 3D beamforming and beamtracking by leveraging the antenna radiation pattern and the steering vectors [30]. For this, $G(\theta_{i,k}(t), \phi_{i,k}(t))$ represents the antenna array gain from AP $k$ to user $i$, which is located at an elevation angle of $\theta_{i,k}(t)$ and azimuth angle of $\phi_{i,k}(t)$ with respect to the APs. The antenna array gain is given by

$$G(\theta_{i,k}(t), \phi_{i,k}(t)) = G_0 \cdot \vec{a}(\theta_{i,k}(t)) \cdot \vec{b}(\phi_{i,k}(t))\,,$$

where $G_0$ represents the constant array gain, while $\vec{a}(\theta_{i,k}(t))$ and $\vec{b}(\phi_{i,k}(t))$ represent the steering vectors in the elevation and azimuth directions, respectively. The vectors $\vec{a}(\theta_{i,k}(t))$ and $\vec{b}(\phi_{i,k}(t))$ are given according to [30] as

$$\vec{a}\,(\theta_i) = \sum_{m=1}^{M} \vec{I}_m \mathrm{e}^{\mathrm{j}(m-1)(kd_z \cos(\theta_i))}, \tag{2}$$

$$\vec{b}\,(\phi_i) = \sum_{n=1}^{N} \vec{I}_n^{\mathsf{T}} \mathrm{e}^{\mathrm{j}(n-1)(kd_y \sin(\theta_i)\sin(\phi_i))}\,, \tag{3}$$

where, $\vec{I}_m$ and $\vec{I}_n$ denote column vectors of ones of sizes $m$ and $n$, respectively. The number of antennas in $z$- and $y$-directions of the antenna array are $M$ and $N$, respectively. The $d_z$ and $d_y$ represent the antenna spacing in the $z$- and $y$-directions, respectively, and $k$ represents the wave number. For ease of reading, we omit the time index $t$ and the superscript $\Gamma$, unless it is necessary to explicitly specify it.

*Definition 2:* For a given clustering strategy $\Gamma$, a power allocation policy $P_{T,ik}$ defines power allocation in each cluster $i$, $\{P_{T,i1}, P_{T,i2}, \ldots, P_{T,iK}\}$, where $P_{T,ik}$ is the allocated power from AP $k$ to AU $i$.

With the above definition, the receive signal-to-interference-plus-noise ratio (SINR) at the target AU $i$ served by cluster $\mathcal{M}_i$ is:

$$\gamma_i^{\mathcal{M}_i} = \frac{\sum_{k=1}^{|\mathcal{M}_i|} h_{ik} P_{T,ik} G(\theta_{i,k}, \phi_{i,k})}{\sigma^2 + \sum_{k=1}^{K} \sum_{\substack{n=1\\ n\neq i}}^{N} h_{ki} P_{T,nk} G(\theta_{n,k}, \phi_{n,k})}, \tag{4}$$

where $\sigma^2$ is the noise power, and the interference power is the sum of all received power from all APs serving on the same resource to other AUs.

### *A. Finite Block-Length Case*

Unmanned aerial vehicle (UAV) applications such as real-time surveillance, emergency response, and industrial inspection require not only seamless service continuity for payload traffic but also ultra-reliable low latency communication (URLLC) for command and control operations. Such communication relies on finite block-length codes, where the decoding error probability is non-zero, thus affecting the reliability. Multi-connectivity has been shown to improve URLLC in such finite block-length regimes [31], [32], [33]. Therefore, our goal is to investigate the impact of cluster reconfiguration and power allocation in this finite block-length regime.

Assuming white Gaussian noise, the channel capacity as a function of the SINR at the AU $i$ is given by:

$$C(\gamma_i^{\mathcal{M}_i}) = \log_2(1+\gamma_i^{\mathcal{M}_i})\,. \tag{5}$$

In the finite block-length regime, there is an approximation for the maximum achievable rate, accounting for the finite sample size, error probability, and coding. The date rate for AU $i$ with block-length $n$ and error probability $\varepsilon$ is given by [34, Eq. (296)]:

$$R_i^*(n,\varepsilon) \approx \mathbb{E}\left\{ C(\gamma_i^{\mathcal{M}_i}) - \sqrt{\frac{V(\gamma_i^{\mathcal{M}_i})}{n}} Q^{-1}(\varepsilon_i) + \frac{\log_2 n}{2n} + \mathcal{O}(1) \right\} \tag{6}$$

where $Q^{-1}$ is the inverse Q-function, which is defined as

$$Q(x) = \frac{1}{\sqrt{2\pi}} \int_x^\infty \exp\left(\frac{t^2}{2}\right) \mathrm{d}t\,.$$

The channel dispersion $V(\gamma_i^{\mathcal{M}_i})$ in (6) is given by

$$V(\gamma_i^{\mathcal{M}_i}) = \left(1 - \frac{1}{(1+\gamma_i^{\mathcal{M}_i})^2}\right)(\log_2 \mathrm{e})^2\,. \tag{7}$$

According to (6), using a finite block length results in a maximum achievable rate that is less than the channel capacity (with an infinite block length). This reduction is caused by the channel dispersion $V(\gamma)$. However, as the block length $n$ becomes very large and the error probability $\varepsilon_i$ approaches zero, the achievable rate can reach the channel capacity.

The minimum decoding error probability (DEP) incurred in the transmission of $b_i$ bits to AU $i$ using a finite block-length of $n$ can be accurately approximated by substituting $R_i = \frac{b_i}{n}$ in (6) and is given by [35]

$$\varepsilon_i(n,b_i) \approx \mathbb{E}\left\{ Q\left( \frac{nC(\gamma_i^{\mathcal{M}_i}) + \frac{1}{2}\log_2 n - b_i}{\sqrt{nV(\gamma_i^{\mathcal{M}_i})}} \right) \right\}. \tag{8}$$

The expectation in (8) is over $\gamma_i^{\mathcal{M}_i}$ since the SINR changes with different channel realizations. In order to have reliable communications, the error probability should be less than a maximum threshold, i.e., $\varepsilon_i(n,b_i) \leq \varepsilon_{\text{max}}$. The maximum DEP constraint can also be written as a SINR constraint as

$$\gamma_i^{\mathcal{M}_i} \geq \gamma_{\text{th}} \approx \exp\left( \frac{Q^{-1}(\varepsilon_{\text{max}})}{\sqrt{n}} + \frac{b_i \ln 2}{n} - \frac{\ln n}{2n} \right) - 1. \tag{9}$$

While we assume that the positions of the AUs and the fading statistics are known, the exact channel state is assumed to be unknown. Hence, the user will be in an outage with a non-zero probability when the SINR at the AU is below a predefined threshold $\gamma_{\text{th}}$, i.e., the outage probability for user $i$ at time $t$ is given as

$$O_i(t) = \Pr\left(\gamma_i^{\mathcal{M}_i}(t) < \gamma_{\text{th}}\right). \tag{10}$$

Depending on the specific use case, there exists an outage probability requirement, denoted as $O_{\text{max}}$, that is deemed acceptable. However, it is essential to acknowledge that this tolerance level is influenced by various factors and may change over time, e.g., when the user moves into a different area.

### B. SINR Outage

In the sequel, we derive an expression for calculating the outage probability from (10) for a single time slot $t$, i.e., for a fixed power allocation and fixed positions of all users. In this case, we can rewrite the outage probability as the probability of a new random variable that comprises of a sum of exponentially distributed random variables with different expected values,

$$\begin{aligned} O_i(t) &= \Pr\left(\gamma_i^{\mathcal{M}_i}(t) < \gamma_{\text{th}}\right) \\ &= \Pr\left(\sum_{k=1}^{|\mathcal{M}_i|} h_{ik} P_{T,ik} G(\theta_{i,k},\phi_{i,k}) < s_i\right) \\ &= \Pr\left(\sum_{k=1}^{|\mathcal{M}_i|} Y_{ik} < s_i\right) \\ &= \Pr\left(T_i < s_i\right) \\ &= 1 - \bar{F}_{T_i}(s_i) \end{aligned} \tag{11}$$

where $s_i = \gamma_{\text{th}}\beta_i$ is the product of the SINR threshold $\gamma_{\text{th}}$ and the interference power $\beta_i$ at user $i$. Based on the Rayleigh fading model, the random variable $T_i$ is given as the sum of exponentially distributed variables $Y_{ik} \sim \text{Exp}(\alpha_{ik})$ with different expected values $\alpha_{ik}$. The expected values are given by the product of transmit power, antenna gain, and path loss. The survival function $\bar{F}_{T_i}$ of $T_i$ is given by [36]

$$\bar{F}_{T_i}(s) = \sum_{k=1}^{K} A_{ik} \cdot \exp\left(-\alpha_{ik} \cdot s\right), \tag{12}$$

$$A_{ik} = \prod_{\substack{j=1 \\ j\neq k}}^{K} \frac{\alpha_{ik}}{\alpha_{ij}+\alpha_{ik}}, \quad \text{for } k=1,\ldots,K. \tag{13}$$

For this expression to hold, we need to assume that all $\alpha_{ik}$ are distinct. However, since they are the product of transmit power, antenna gain, and path loss, this assumption will hold almost surely in practice.

### C. Problem Formulation

The joint dynamic clustering and power allocation for APs is of utmost importance to ensure both reliable and energy-efficient communication. To accomplish this objective, we present an optimization problem aimed at finding the optimal serving cluster that satisfy the QoS requirements for each user and the corresponding power allocation vector.

Let $\boldsymbol{\mathcal{M}} = \{\mathcal{M}_i, \forall i \in \mathcal{N}\}$ denote the variable for AP clustering. Let $\boldsymbol{\mathcal{P}} = \{P_{T,ik}, \forall i \in \mathcal{N}, \forall k \in \mathcal{K}\}$ denote the transmit power variable for the multi-connectivity users. Based on this, we use a scalarization to formulate the general multi-objective optimization problem of APs clustering and power allocation as

$$\max_{\boldsymbol{\mathcal{P}},\boldsymbol{\mathcal{M}}} \frac{1}{N}\sum_{i=1}^{N}\mathbb{1}\left(\mathcal{M}_i(t)=\mathcal{M}_i(t-1)\right)+\frac{\sum_{i=1}^{N} b_i}{nP_{\text{total}}/(1-\varepsilon_{\max})} \tag{14a}$$

$$\text{s.t. } C_1 : \varepsilon_i \leq \varepsilon_{\max},\ \forall i \tag{14b}$$

$$C_2 : P_{T,ik} \leq P_{\max}, \quad \forall i,k,t \tag{14c}$$

$$C_3 : |\mathcal{M}_i(t)| \geq 1, \quad \forall i,t \tag{14d}$$

$$C_4 : \gamma_i^{\mathcal{M}_i}(t) \geq \gamma_{\text{th}}, \quad \forall i,t \tag{14e}$$

where $\varepsilon_{\max}$ is the maximum error threshold tolerable for the reliable communication, $b_i$ is the number of information bits transmitted to AU $i$, $n$ is the block-length, and $P_{\text{total}} = \sum_{k=1}^{K}\sum_{i=1}^{N} P_{T,ik}$ is total transmitted power. The first term in (14a) aims to maintain the stability of the serving clusters for the AUs moving within the area, while the second term seeks to maximize the number of bits successfully transmitted using the minimum total transmit power. It can be observed that given $\varepsilon_{\max}$, $n$, and $b_i$ are constant, maximizing the successfully transmitted bits is equivalent to minimizing the total power consumption from all APs.

### D. Problem Decomposition

As shown in [1], the original problem remains scalable when the number of APs in the service area is limited. However, as the number of APs increases, the action space grows significantly, making it challenging to find a solution for the joint objective function within a finite time. To address this challenge, we decompose the optimization problem in (14a) into two subproblems, separating the tasks of cluster reconfiguration and power allocation.

*1) Subproblem 1: Optimal Cluster Reconfiguration:* The first subproblem aims to maintain the stability of the serving clusters, which is formulated as:

$$\Gamma : \max_{\boldsymbol{\mathcal{M}}} \frac{1}{N}\sum_{i=1}^{N}\mathbb{1}\Big(\mathcal{M}_i(t)=\mathcal{M}_i(t-1)\Big) \tag{15a}$$

$$\text{s.t. } C_1 - C_4. \tag{15b}$$

The problem poses a combinatorial challenge and is difficult to solve due to the non-convex nature of its objective function. Typically, there is no computationally efficient or systematic method for achieving an optimal solution to this type of problem. However, it is well-suited for an iterative RL approach. To ensure this subproblem remains bounded and converges in time, we modify the problem by incorporating constraint $C_4$ into the objective function. Further details are given in Section IV-A.

*2) Subproblem 2: Optimal Power Allocation:* Consider a given fixed clustering policy $\pi$, the optimal power allocation policy $\boldsymbol{\mathcal{P}}^{\star}_{i,k}$ can be obtained as a solution to the following optimization problem for every $t$:

$$\boldsymbol{\mathcal{P}}^{\star}_{i,k} : \max_{\boldsymbol{\mathcal{P}}} \frac{\sum_{i=1}^{N} b_i}{nP_{\text{total}}/(1-\varepsilon_{\max})} \tag{16a}$$

$$\text{s.t. } C_1 - C_4 \tag{16b}$$

The objective in (16a) is to maximize the number of bits transmitted while satisfying reliability constraints, with the least possible total transmitted power. The optimal power allocation $\boldsymbol{\mathcal{P}}^{\star}_{i,k}$ minimizes the transmitted power in the link between user $i$ and the serving AP cluster for all user-AP links at each time $t$. Thus, $\boldsymbol{\mathcal{P}}^{\star}_{i,k}$ provides the optimal power allocation for any given clustering $\Gamma$.

### E. Mobility Model

In this work, we employ a realistic and tractable mobility model to capture the mobility of UAVs. In particular, we use the model provided in [28] using coupled stochastic differential equations. By utilizing estimated positions instead of actual ones, the model incorporates more realistic device trajectories and considers imperfect navigation. The advantage of this approach lies in its ability to generate smoother and realistic trajectories. Additionally, it provides better control over velocity through correlation parameters, which influence stability and mobility based on distance-velocity relationships. Additional variance parameters scale Brownian perturbations, offering flexibility in introducing stochastic variations. For detailed explanations of the model, please refer to the original work [28].

### F. User Handling

In a multi-connectivity wireless interference network, the dynamic clustering of APs depends on the number of users being served by the network. The arrival and departure of users within the network's coverage area influence the clustering decisions. Therefore, for a more realistic and dynamic scenario, our system needs to efficiently manage users entering and leaving the coverage area without disrupting the already established clusters.

*1) Users Entering the Coverage Area:* When a completely new mobile user enters the coverage area of the system, they were not previously associated with the system. As a result, they become part of the active user group and are served by a new AP cluster. The formation of a new serving clusters without affecting the other clusters ensures that new users seamlessly receive services within the existing system framework.

*2) User Leaves the Coverage Area:* On the other hand, active users can become inactive, e.g., by moving out of the service area or switching off their device. As a result, they transition from an active to a non-active state and are no longer served by the AP cluster they were previously associated with. The freed APs can then be used to become part of the serving clusters of the active users. This allows to efficient management of resources and ensure optimal service distribution to active users.

## IV. H-MADRL For Dynamic Clustering and Power Allocation

To address the joint problem of dynamic clustering and power allocation in a scalable and adaptive manner, we adopt a hierarchical multi-agent deep reinforcement learning (H-MADRL) framework. This approach builds on the decomposition in Section III-D, where the original optimization problem is separated into two interdependent subproblems: cluster reconfiguration and power control. These subproblems exhibit sequential and spatial dependencies, making them well-suited for a learning-based formulation.

We model the overall decision-making problem as a partially observable Markov decision process (POMDP), defined by the tuple $(\mathcal{S}, \mathcal{A}, \mathcal{P}, r, \gamma)$. The state space $\mathcal{S}$ captures the time-varying network conditions relevant to clustering and power allocation. The joint action space $\mathcal{A}$ consists of high-level clustering decisions made at the edge cloud and low-level power allocations performed by distributed AP agents. The transition function $\mathcal{P}$ evolves based on user mobility model [28], the wireless channel evolution including height-dependent LoS/NLoS and fading [29], and the interference and reliability changes caused by new clustering and power decisions. The reward function $r(\mathbf{s}_t, \mathbf{a}_t)$ promotes energy efficiency, cluster stability, and communication reliability. The objective is to learn a joint policy that maximizes the expected cumulative discounted reward, $\mathbb{E}[\sum_t \gamma^t r(\mathbf{s}_t, \mathbf{a}_t)]$, where the action $\mathbf{a}_t$ combines the high-level clustering decision and distributed low-level power allocations.

To solve this POMDP, we propose a H-MADRL framework as shown in Figure 1. In this framework, the edge cloud first determines the high-level action, specifically the clustering strategy $\Gamma$ for each user $i \in \mathcal{N}$. Subsequently, each low-level AP $k \in \mathcal{K}$ updates its power allocation policy $P_{T,ik}$ for each assigned user. This update is based on the assigned users $N_k$, their locations, and channel conditions. The interactions among the low-level agents collectively determine the system state for the high-level agent, prompting the edge cloud to revise its clustering strategy in the subsequent decision epoch. The hierarchical structure decomposes the complex joint problem into tractable sub-policies, effectively reducing the search space for each agent and improving learning efficiency. The next subsections describe this hierarchical framework in detail, including the structure of observation and action spaces and the learning objectives for each agent level.

## A. High-Level: Dynamic Cluster Reconfiguration

First, we discuss the high-level agent in detail, in particular its observation space, action space, and reward function.

*1) High-Level Observation Space:* The observation space of the high-level agent, located at the edge cloud, must include network-level information about the connected APs and AUs for optimal clustering. The high-level observation space includes the location information of the AUs $\boldsymbol{x}, \boldsymbol{y}, \boldsymbol{z} = \{x_i, y_i, z_i \quad \forall i\}$, user load of the APs $\boldsymbol{\mathcal{N}} = \{\mathcal{N}_k, \forall k\}$, the channel condition between connected APs and the AUs $\boldsymbol{h} = \{h_{i,k}, \forall i, k\}$, and the current clustering $\boldsymbol{\mathcal{M}} = \{\mathcal{M}_{i,k}, \forall i, k\}$. Consequently, the observation space $\bar{S}_t \in \mathcal{S}$ of the high-level agent is expressed as:

$$\bar{S}_t \triangleq \{\boldsymbol{x}(t), \boldsymbol{y}(t), \boldsymbol{z}(t), \boldsymbol{l}(t), \boldsymbol{h}(t), \boldsymbol{\mathcal{M}}(t-1)\} \tag{17}$$

with a dimensionality of $3N + K + 2NK$.

*2) High-Level Action Space:* At each time step, the main role of the high-level agent is to take an action $\bar{a}_t \triangleq [\mathcal{M}_1, \mathcal{M}_2, \ldots, \mathcal{M}_N]$, which assigns the serving cluster $\mathcal{M}_i$ to each AU $i \in \mathcal{N}$. This clustering action $\bar{a}_t \in \mathcal{A}$ constitutes the first component of the joint action, and results in a high-level action space of dimensionality $NK$.

*3) High-Level Reward Function:* The reward function for the high-level agent guides its policy $\pi_\theta$ parameterized by $\theta$ by rewarding stability in cluster configurations, i.e., by minimizing the number of cluster reconfigurations, as described in (15a). To connect the high-level policy with the low-level actions of the APs, the reward is structured around two key objectives: maintaining stable serving clusters for the AUs and reducing the number of AUs experiencing SINR outages. Thus, the agent is rewarded when clusters remain stable, provided that users are not in outage. The instantaneous reward $\bar{r}$ that the agent receives for a given action in a particular state is defined as:

$$\bar{r} = \frac{\omega_1}{N} \sum_{i=1}^{N} \mathbb{1}\Big(\mathcal{M}_i(t) = \mathcal{M}_i(t-1)\Big) - \frac{\omega_2}{N} \sum_{i=1}^{N} \mathbb{1}(O_i > O_{\max}), \tag{18}$$

where the non-negative weights $\omega_i$, $i \in \{1, 2\}$, are used to balance between the individual objectives. The reward $\bar{r}$ in (18) increases when the number of stable clusters or non-outage users increases. The $O_i$ is obtained from (10) and depends on the SINR threshold.

## B. Low-Level: Multi-Agent Optimal Power Allocation

Through the action $\bar{a}_t$ of the learned policy $\pi_\theta$, the edge cloud assigns a serving cluster $\mathcal{M}_i$ for each AU $i \in \mathcal{N}$. Consequently, an AP $k \in \mathcal{K}$ is allocated a set of users $\mathcal{N}_k \subseteq \mathcal{N}$ to be served by this AP. The APs, using a multi-agent learned policy, optimize the power allocation for their assigned users. To facilitate this, we outline the basic RL components for the low-level multi-agents in the following.

*1) Low-Level Observation Space:* The low-level observation space $s_t^k$ is designed to encompass both local information pertinent to the agent and shared data from the higher level $s_t^k = \{\bar{s}_t^k, \bar{a}_t^k\}$. Specifically, the observation space for the $k$-th agent includes the assigned users $N_k \subseteq \mathcal{N}$, their location information $\boldsymbol{x}, \boldsymbol{y}, \boldsymbol{z} = \{x_j, y_j, z_j : j \in N_k\}$, the line-of-sight (LoS) conditions for assigned users $LoS_{k,j}, j \in N_k$, and a set detailing the assigned clusters from the higher level $\boldsymbol{\mathcal{M}}_k = \{\mathcal{M}_j, \forall j \in N_k\}$. Consequently, the observation space $S_t^k \in \mathcal{S}$ of the low-level $k$-th agent is expressed as:

$$S_t^k \triangleq \{\boldsymbol{x}(t), \boldsymbol{y}(t), \boldsymbol{z}(t), LoS_{j,k}, \boldsymbol{\mathcal{M}}_k(t)\} \tag{19}$$

with a dimensionality of $4N_k + N_k K$.

*2) Low-Level Action Space:* The low-level agents at APs operate independently, utilizing a learned policy. At each time epoch $t$, the higher-level agent performs clustering actions for each user within the service area, and each low-level agent $k$ takes the action $a_t^k$, which determines the optimal power allocation $P_{j,k}^{\star}, \forall j \in \mathcal{N}_k$. An agent $k$ develops a policy $\pi_{\theta_k}$ parameterized by $\theta_k$ that exploits the dynamic characteristics of the channel and the clustering from the higher level to optimize power allocations from various APs in the cluster. The policy's primary objective is to maximize the objective function (16a) while adhering to constraints. The action space of low-level agent $k$ is expressed as:

$$a_t^k \triangleq [P_{1,k}^\star, P_{2,k}^\star, \ldots, P_{N_k,k}^\star] \tag{20}$$

This power allocation action $a_t^k \in \mathcal{A}$ constitutes the second component of the joint action in $\mathcal{A}$, with dimensionality $N_k$ for agent $k$.

*3) Low-Level Reward Function:* The reward for the low-level agent aims to increase the fraction of users associated to it, while using minimum power and satisfying the constraints. In particular, the instantaneous reward is expressed as a continuous function that seeks to minimize the total transmitted power while penalizing the violation of constraints. The low-level reward function $r_t^k$ for agent $k$ at time $t$ is given as:

$$r_t^k = \left(1 - \omega_1 \frac{\sum_j^{N_k} P_{j,k}^\star}{N_k P_{T\max}}\right) - \frac{\omega_2}{N_k} \sum_{i=1}^{N_k} \mathbb{1}(\varepsilon_i > \varepsilon_{\max}) \tag{21}$$

where the non-negative weights $\omega_i$, $i \in \{1, 2\}$, are used to balance between the individual objectives. The reward $r$ in (21) increases as the total transmit power decreases and penalizes the agent when any assigned user's instantaneous error probability $\varepsilon_i$ exceeds the predefined reliability threshold $\varepsilon_{\max}$. Towards this goal, we implement the multi-agent proximal policy optimization (MAPPO) algorithm to derive the power allocation policy $P_{j,k}^\star$.

### C. The Proposed Hierarchical MAPPO Algorithm

To enable the H-MADRL framework, we use a single agent proximal policy optimization (PPO) algorithm at the edge cloud for high-level clustering decisions and MAPPO at the APs for low-level power control. We refer to the resulting algorithm as H-MAPPO to emphasize the use of PPO within the hierarchical multi-agent design and pseudocode is provided in Algorithm 1. In the first episode, the high-level edge cloud agent selects the clustering of the APs. This clustering decision then updates the observation space for the low-level APs agents, allowing each AP to determine the power allocation based on the local observations, as detailed in lines 7 through 9 of Algorithm 1. While this decomposition may lead to some performance trade-offs compared to fully joint optimization, it provides significant benefits in terms of scalability and modularity. The use of PPO at the high level and MAPPO at the low level enables stable policy updates in large and partially observable action spaces.

To improve coordination between levels and sample efficiency under partial observability, we employ an action–observation transition mechanism in which the high-level clustering action is injected into the observation space of the low-level AP agents before power control as shown in Figure 2.

To enable practical deployment, the proposed framework assumes the policies are trained entirely in a digital-twin environment using on-policy rollouts, and then executed as frozen networks. During deployment, the edge cloud emits clustering actions over the optical backhaul, APs run local inference for power control using their augmented observations, and no online exploration is performed.

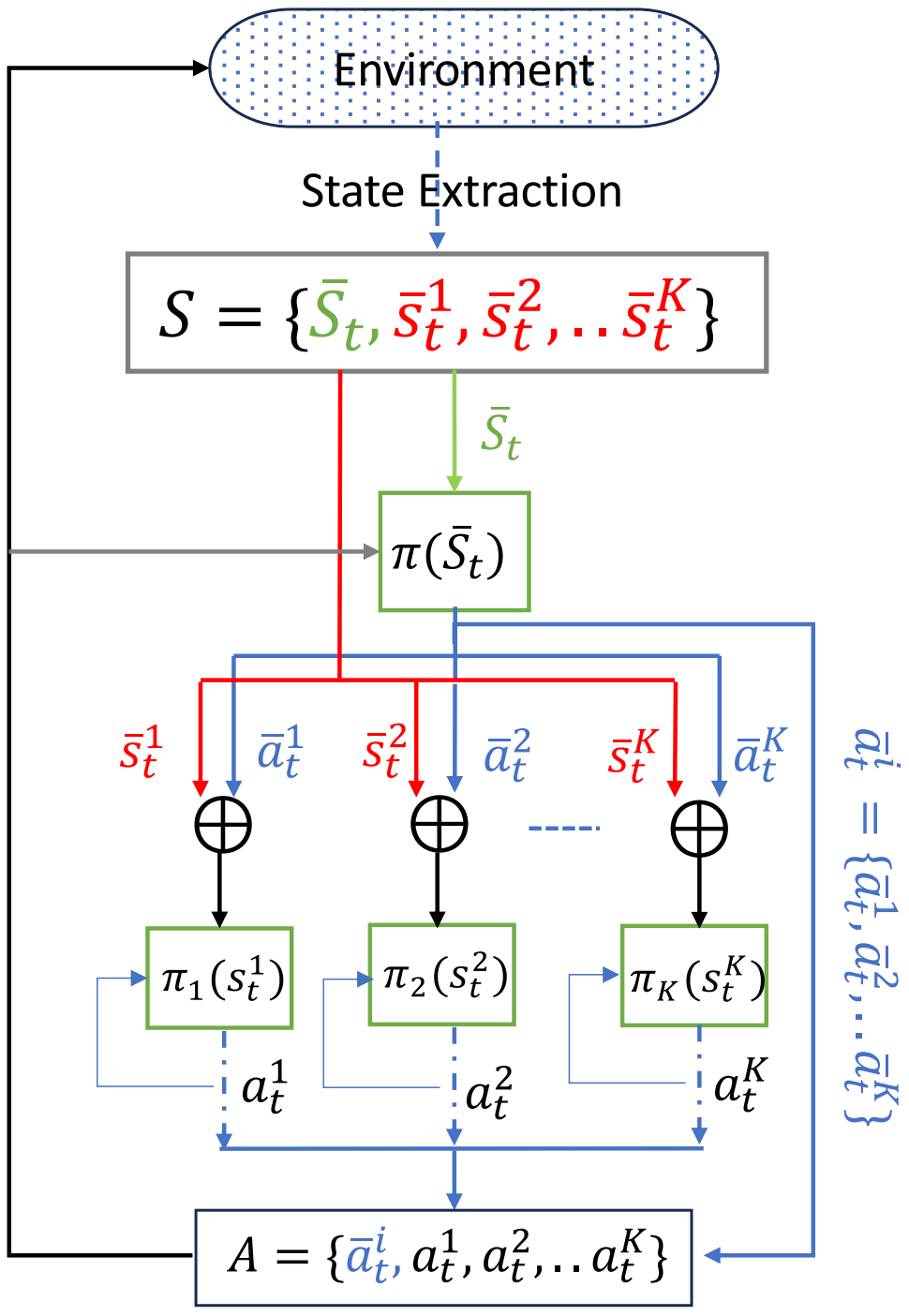

Fig. 2. Proposed action-observation transition-driven hierarchical multi-agent proximal policy optimization (H-MAPPO) framework.

### D. MAPPO Policy Training

In the MAPPO algorithm, each policy $\pi_{\theta_k}$ parameterized by $\theta_k$, is a function that maps states to actions. The policy is typically represented by a neural network, which takes the state as input and outputs a probability distribution over the possible actions. Each policy is trained independently, meaning that the data collected from one policy is not used to train the other policies. However, the policies can share the same architecture and weights, and can be trained in parallel. Multi-agent proximal policy optimization (MAPPO) uses the same core principles as single-agent PPO, with modifications to handle multiple agents and their interactions. Derived from the Trust Region Policy Optimization algorithm [37], [38], PPO introduces a clipped surrogate objective to enhance stability in policy updates. Each agent updates its policy to maximize the expected reward, using the following clipped surrogate objective [39]:

$$L_k^{\text{CLIP}}(\theta_k) = \mathbb{E}_t\left[\min\left\{\rho_t(\theta_k)\hat{A}_t^k, \text{clip}\left(\rho_t(\theta_k), 1-\eta, 1+\eta\right)\hat{A}_t^k\right\}\right] \tag{22}$$

where $\rho_t(\theta_k)$ is the probability ratio between the new and old policies for the action $a_t^k$:

$$\rho_t(\theta_k) = \frac{\pi_{\theta_k}(a_t^k | s_t^k)}{\pi_{\theta_k^{\text{old}}}(a_t^k | s_t^k)} \tag{23}$$

and $\hat{A}_t^k$ is the advantage function estimate for agent $k$. The clipping parameter $\eta$ controls the extent of policy updates to

**Algorithm 1** H-MAPPO Based Dynamic Clustering and Power Allocation

**Input:** $K$, $N$, $\varepsilon_{\max}$, $P_{\max}$, and $\gamma_i^{th}$
**Output:** Clustering strategy $\Gamma$, Optimal power allocation $P_{i,k}^{\star}$

**for** each episode $\leftarrow 1$ to end **do**
  Initialize: High level observation space $\bar{S}_t$, low-level observation space for each agent $S_t^i$
  **while** not done **do**
    **if** first episode **then**
      Only high-level agent accepts observation space and returns the action space
    **end if**
    $\bar{A}_t = \{\bar{a}_t^1, \bar{a}_t^2, \ldots, \bar{a}_t^K\}$: Action space from trained high level agent
    $S_t^i = \{\bar{a}_t^i, x_i(t), l_i(t), LoS_i\}$: Updated observation space of a low-level agent $i$
    $A_t = \{a_t^1, a_t^2, \ldots, a_t^K\}$: Action space from trained low-level agents
    Calculate the error probability from (8) using $\{\bar{A}_t, A_t\}$
    Obtain the high level reward using (18) and low-level reward using (21)
    Update the status of the users
  **end while**
**end for**

ensure stability. Advantage estimation quantifies how much better an action is compared to the expected value. MAPPO employs Generalized Advantage Estimation (GAE) to reduce variance and improve stability:

$$\hat{A}_t^k = \sum_{l=0}^{\infty} (\gamma\lambda)^l \delta_{t+l}^k, \tag{24}$$

where $\delta_t^k$ is the temporal difference error for agent $k$:

$$\delta_t^k = r_t^k + \gamma V(s_{t+1}^k) - V(s_t^k). \tag{25}$$

Here, $\gamma$ is the discount factor, $\lambda$ is the GAE parameter, and $V(s_k)$ is the state value function for agent $k$. In MAPPO, each agent learns its policy independently, optimizing its objective based on individual experiences:

$$\theta_k \leftarrow \theta_k + \alpha \nabla_{\theta_i} L_k^{\text{CLIP}}(\theta_k), \tag{26}$$

where $\alpha$ is the learning rate. Agents interact within the environment, which provides feedback in the form of rewards that guide the policy updates.

### *E. Deployment Scenario*

We formulate our problem as a POMDP within OpenAI's Gym environment. Our hierarchical multi-agent framework is implemented using Ray's RLlib library [40]. RLlib offers a scalable, flexible, and efficient solution for training RL models across multiple nodes and GPUs. By harnessing Ray's distributed computing capabilities, RLlib parallelizes simulations, experience collection, and model updates, significantly accelerating training processes and enabling the handling of large-scale RL tasks. The initial agent hyperparameters are summarized in Table III, having been empirically determined through multiple iterations. The source code of our implementation for reproducing all of the shown results will be made publicly available with the final version of the paper.

TABLE III
HYPERPARAMETERS EMPLOYED FOR TUNING OUR MODEL

| Parameters | Value | Parameters | Value |
|---|---|---|---|
| Learning rate | $10^{-5}$ | Discount factor ($\gamma$) | 0.99 |
| Batch size | 4000 | GAE parameter ($\lambda$) | 1.0 |
| Iterations | $2 \cdot 10^6$ | PPO Clip parameter ($\eta$) | 0.3 |

In practice, the digital twin can be calibrated with limited field telemetry (e.g., SINR traces) and light domain randomization, and retraining policies offline, and then redeploying them as frozen policies. This workflow leverages real data to refine environment parameters and robustness, while avoiding online exploration in the live network.

## V. NUMERICAL EVALUATION

In this section, we showcase the effectiveness of our proposed H-MAPPO implementation for solving (14a). Additionally, we benchmark our `H-MAPPO` algorithm against the following baseline methods:

- `MSAC`: A centralized off-policy approach called masked soft actor-critic (MSAC) based clustering and power allocation from [1].
- `MAPPO`: An on-line and distributed approach without any hierarchical policy division-based clustering and power allocation, where each agent decides to join the serving cluster for a user and allocated power.
- `Opportunistic`: An opportunistic clustering approach from [16], where a central entity opportunistically decides to include an O-RU (AP) in a cluster for user service based on channel gains.
- `JCPC`: A joint clustering and power control (JCPC) scheme proposed in [8], employs iterative algorithms based on penalty convex-concave procedures and successive convex approximation to jointly optimize AP clustering, and power allocation.
- `Closest`: A simple approach in which only the nearest O-RU (AP) serves a user with maximum power.

For the numerical evaluation, there are $N = 6$ users in a square area of $3\,\text{km} \times 3\,\text{km}$. They are served by $K = 19$ APs, which are placed randomly within this area at a height of $25\,\text{m}$. Each AP is equipped with $L = 16$ antennas. The model of the path-loss follows [29], where we set the carrier frequency to $2.0\,\text{GHz}$ to accommodate a broader range for command and control traffic for the UAVs. The noise power at the receiver is given as $\sigma^2 = N_0 B$, where $B = 10\,\text{MHz}$ is the communication bandwidth, and $N_0 = -174\,\text{dBm/Hz}$ is the noise spectral density. The UAVs move randomly following the mobility model described in Section III-E across the area. For a fair comparison, we keep all parameters of the communication system the same for all algorithms.

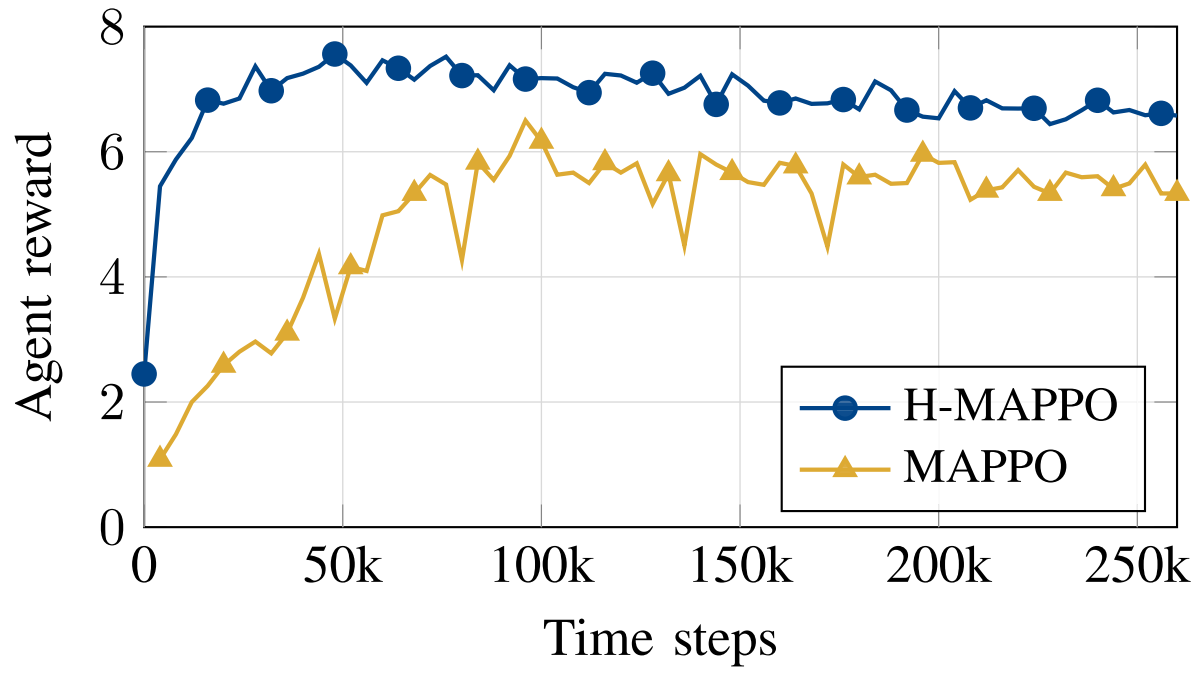

Fig. 3. Numerical results comparing the rewards obtained during training iterations for both H-MAPPO and MAPPO.

### *A. Learning Efficiency and Complexity of H-MAPPO*

We evaluate the learning efficiency of the proposed `H-MAPPO` algorithm against the distributed `MAPPO` baseline. While `MAPPO` requires each AP to jointly learn both clustering and power allocation policies, `H-MAPPO` separates these tasks into a high-level agent responsible for clustering and low-level agents for power control. This hierarchical decomposition reduces the size and coupling of the action space at each level, simplifying policy learning and improving credit assignment.

As shown in Figure 3, `H-MAPPO` converges more quickly and achieves higher cumulative rewards, outperforming `MAPPO` by approximately $15\,\%$. The improvement is particularly evident in the early stages of training, where disentangled decision-making allows faster exploration and policy stabilization. Moreover, by allowing each AP to optimize power allocation independently based on localized observations, the system adapts better to dynamic user presence and channel variability.

In terms of computational complexity, `H-MAPPO` and `MAPPO` have comparable theoretical bounds. However, as summarized in Table IV, `H-MAPPO` benefits from parallel low-level training and reduced per-agent policy complexity due to task decoupling. In contrast, centralized methods like `MSAC` and `JCPC` rely on a single decision-maker, which makes them harder to scale as the number of users or APs increases. Moreover, while `JCPC` is theoretically optimal, it has high-order polynomial complexity and is infeasible even for small system sizes. Opportunistic clustering offers the fastest runtime but lacks adaptability and learning capability. The hierarchical design of `H-MAPPO` offers a practical compromise between coordination and efficiency, enabling scalable and fast learning without sacrificing decision quality.

### *B. Reliability Performance*

In Figure 4, we demonstrate the reliability performance of our proposed action-observation transition-driven `H-MAPPO` algorithm. One of the primary objectives of our problem is to ensure reliable data transmission with a DEP below a given threshold, i.e., $\varepsilon_i \leq \varepsilon_{\max}$. The results for DEP threshold violations are presented in Figure 4(a) and the distribution of the SINR outage probability $O$ in Figure 4(b). Although these two objectives are interdependent, in our proposed action-observation transition-driven `H-MAPPO` algorithm, the DEP threshold constraint governs the low-level policy, while the SINR outage constraint influences the high-level policy.

In Figure 4(a), we observe the results of the average DEP threshold violations experienced by the users under different schemes. The proposed `H-MAPPO` algorithm and the centralized `MSAC` scheme exhibit performance comparable to the optimization-based clustering approach `JCPC` for medium to high DEP thresholds. In contrast, the fully distributed `MAPPO` suffers from higher violation rates due to its lack of coordination. The `Opportunistic` and `Closest` schemes, which do not incorporate any reliability optimization, experience violation probabilities close to one, indicating their inability to ensure reliable transmission across different network conditions. Furthermore, as the DEP threshold decreases, the number of outages increases. This is expected, as maintaining a low error probability in a dynamic environment becomes challenging due to fluctuating channel conditions.

Next, in Figure 4(b), we present the SINR outage probability results. The cumulative distribution function (CDF) illustrates the distribution of outage probabilities $O$ experienced by users as they move through the service area. As can be seen from the figure, the proposed `H-MAPPO` algorithm achieves performance comparable to the centralized `JCPC` scheme. Specifically, the cumulative distribution function (CDF) for `H-MAPPO` approaches one when the outage probability is around $10^{-8}$, indicating that the user's SINR outage probability almost never exceeds this threshold. For `JCPC`, the corresponding value is approximately $10^{-9}$, an order of magnitude lower than that of `H-MAPPO`. In contrast, the centralized `MSAC` scheme reaches this point around $10^{-5}$, which is three orders of magnitude higher than the proposed algorithm. Nonetheless, both `H-MAPPO` and `JCPC` significantly outperform the baseline schemes, namely `MAPPO`, `Opportunistic`, and `Closest`. The superior performance of `H-MAPPO` stems from its ability to adapt to strict reliability constraints through dynamic power control by low-level agents, guided by the maximum error threshold. By considering user positions, LoS conditions, and spatial relationships between neighboring APs, the algorithm minimizes interference and ensures reliability.

Although the `Opportunistic` strategy performs better than the `Closest` approach, the latter reduces interference by limiting the number of serving APs per user. However, this reduction in interference comes at the cost of reduced received power, which increases the likelihood of outages. Both strategies fall short of optimizing for the stricter reliability demands of the system. These results highlight the effectiveness of our hierarchical framework, which combines the reliability of a centralized approach with the scalability and efficiency of a distributed approach.

### *C. Transmit Power Performance*

The second objective of our problem is to minimize the total transmitted power from the serving clusters while ensuring the system meets its reliability performance targets. Achieving this with minimal power is crucial for system energy efficiency. The comparison of the total transmitted power across different algorithms is shown in Figure 5. Since the transmit power in

TABLE IV
COMPARISON OF COMPUTATIONAL COMPLEXITY

| Method | Per-Step Comp. Complexity | Execution | Runtime per Step (ms) |
|---|---|---|---|
| MSAC | $\mathcal{O}(2LH^2 + NK)$ | Central | 6–10 |
| MAPPO | $\mathcal{O}(KLH^2 + NH)$ | Parallel | 2–4 |
| Opportunistic | $\mathcal{O}(N \cdot K)$ | Central | 0.5–2 |
| JCPC | $\mathcal{O}(N^7 K^{3.5})$ | Central | ≫3000 |
| Proposed H-MAPPO | $\mathcal{O}(KLH^2 + NKH)$ | Parallel (low-level) | 4–8 |

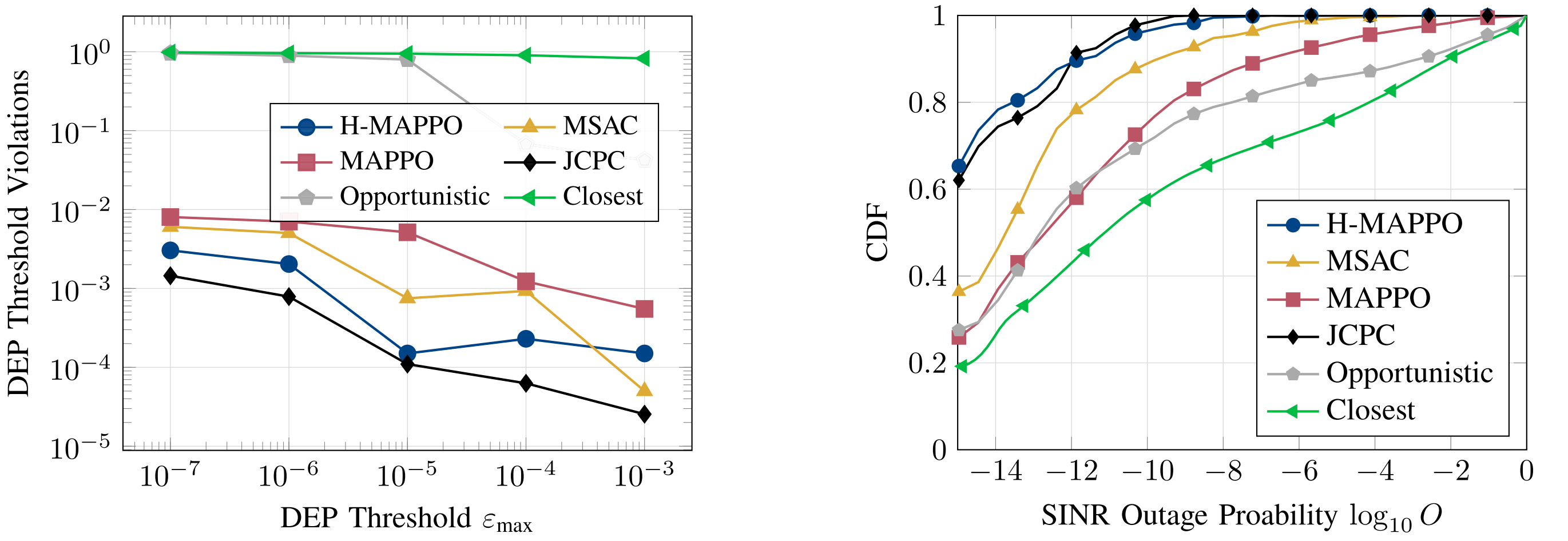

(a) DEP threshold violations experienced by the UAVs for different DEP thresholds $\varepsilon_{\max}$.

(b) CDF of the outage probability $O$ experienced by the UAVs in service area.

Fig. 4. Numerical results of the DEP threshold violation and the SINR outage probability $O$. (Section V-B).

the Closest scheme is constant, it has been omitted from the graph.

As shown in Figure 5(a), the proposed H-MAPPO algorithm achieves the DEP threshold with the lowest transmit power. In contrast, the centralized JCPC and MSAC schemes require the highest transmit power. The slight performance advantage of JCPC in terms of outage probability in Figure 4 comes at the expense of higher power consumption, indicating that JCPC meets the DEP thresholds by expending extra power. Although MSAC also consumes more power, it delivers better reliability than MAPPO, which suffers from a higher outage rate. Overall, the proposed H-MAPPO strikes the best balance, matching the reliability performance with the highest power efficiency. Additionally, it is observed that as the DEP requirement becomes stricter, the total transmitted power increases to boost the SINR and reduce transmission errors.

Meanwhile, the Opportunistic scheme, which is not optimized for meeting reliability constraints, uses on an average over $90\,\%$ of the available power and is unaffected by the DEP thresholds. The Closest scheme operates with a single AP at maximum power, resulting in a constant graph at $1/K$, i.e., $1/19 \approx 0.05$ for this particular example. Therefore, this scheme is not shown in Figure 5(a).

The observations from Figure 5(b) reveal that the learning-based H-MAPPO, MSAC, and MAPPO and optimization based JCPC cluster and power allocation schemes use less than $60\,\%$ of the maximum available transmit power $90\,\%$ of the time. While non-learning Opportunistic scheme uses less than $60\,\%$ of the maximum available transmit power only $30\,\%$ of the time. This highlights that these learning-based methods not only excel at mitigating outages and meeting stringent reliability requirements, but they also achieve these goals with significantly reduced total transmitted power.

### *D. Clustering Size and Scalability Performance*

The third objective in our problem is the reconfiguration of the serving clusters as the users move. The distribution of the average cluster size for serving mobile UAVs is depicted in Figure 6(a). Notably, the distribution for H-MAPPO is centered around smaller cluster sizes, indicating that the high-level clustering decision typically involves half or fewer APs per serving cluster. Having a smaller number of APs serving a user simplifies the coordination and is therefore beneficial. Through training, the high-level agent in H-MAPPO effectively balances SINR outage requirements and the number of APs in each cluster. In particular, almost no clusters with more than 11 APs have been formed by the H-MAPPO algorithm. The cluster size distribution of the H-MAPPO reflects that the scheme uses less than half of the APs for forming clusters. A similar shape can be observed for the distribution of the cluster size using the JCPC algorithm. It has an even stronger emphasis on small cluster sizes and more than half of the found clusters are of size 4 or smaller. For the other algorithms, the distribution significantly shifts to higher cluster sizes. The distribution for MAPPO shows that it uses any size of more than 11 APs with similar probability of around $5\,\%$. The H-MAPPO shows the highest probability for a cluster size of 7, whereas

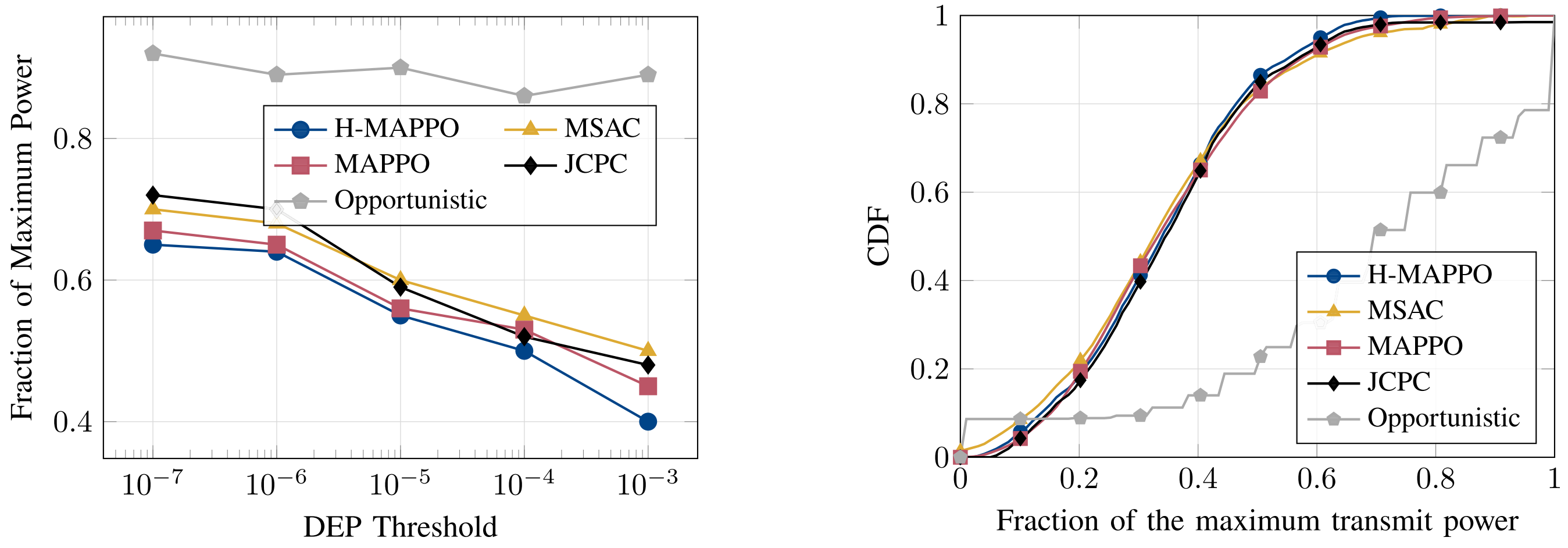

(a) Fraction of total transmitted power for different DEP thresholds $\varepsilon_{\max}$.

(b) CDF of the total transmit power.

Fig. 5. Numerical results of fraction of maximum power utilized and the distribution of the total transmit power of the system normalized by the maximum available power (Section V-C).

for MSAC and MAPPO, the peak occurs at 8. The variation in cluster sizes is due to the separation of clustering and power allocation in H−MAPPO, where power is optimized for a fixed cluster size. In contrast, while MSAC centralizes the power allocation and clustering decisions, directly influencing cluster sizes, MAPPO fully decentralizes these decisions, resulting in a wider distribution of cluster sizes.

Meanwhile, the Opportunistic scheme, driven by favorable channel conditions and full power transmission, often involves more than 50 % of APs in clusters, leading to excessive power usage. The Closest algorithm uses only single AP for connectivity, its results are thus not shown in Figure 6(a). Additionally, it is noteworthy that both H−MAPPO and MSAC exhibit near-zero probability for cluster sizes greater than or equal to 13, whereas the Opportunistic scheme peaks at a cluster size of 16, essentially involving all APs in the cluster.

The results for the cluster reconfiguration of the serving clusters for the considered clustering approaches are shown in Figure 6(b). The proposed H−MAPPO scheme outperforms all the others with a cluster reconfiguration rate of only 9 %, whereas the optimization-based JCPC approach undergoes frequent changes with a high reconfiguration rate of 83 %. Overall, the learning-based clustering schemes demonstrate significantly better performance compared to non-learning-based schemes in terms of cluster stability. This is because a learning agent can implicitly capture long-term spatiotemporal patterns of user mobility and network dynamics, enabling it to anticipate future states and adaptively form more stable serving clusters. In contrast, optimization-based or opportunistic strategies often make decisions based solely on instantaneous metrics, leading to frequent reconfigurations.

To highlight the necessity of a hierarchical approach over centralized or fully distributed methods for joint clustering and power allocation, we evaluate the scalability of our proposed H-MAPPO algorithm against the centralized MSAC method and the fully distributed MAPPO baseline. The results in Figure 7 show the time required to complete an episode in H-MAPPO, MAPPO, and MSAC, relative to the maximum episode time, as the number of APs in the network increases. The findings highlight the superior scalability of the H-MAPPO algorithm, where the decision-making for clustering and power allocation is distributed across two levels: the high-level edge cloud for clustering and the low-level APs for power control. In contrast, MSAC relies on a single centralized agent to manage all decisions, making it less scalable as the network grows. Although MAPPO shows the lowest episode time for small network sizes due to its lightweight distributed agents, its training time increases more rapidly than H−MAPPO as the number of agents grows. This is because each MAPPO agent learns both clustering and power control in a fully distributed setting, leading to higher coordination complexity and slower convergence in larger networks.

A key observation from the results is that when the number of APs doubles from 16 to 32, the training time for H−MAPPO increases by only about 10 % per episode, while the same increase in MSAC leads to a 90 % rise in training time. For MAPPO, the increase is around 70 %, confirming that while distributed, the absence of hierarchical decomposition impacts its scalability. This stark difference underscores the efficiency of H−MAPPO, as it distributes the decision-making process across agents, reducing the computational load on a single entity. Moreover, this improvement in scalability does not come at the cost of performance. The distributed nature of H-MAPPO allows it to maintain or even improve system performance while scaling more efficiently than the centralized MSAC approach.

*Limitations:* While H−MAPPO demonstrates improved performance and learning efficiency, it introduces additional coordination complexity between the high-level and low-level agents. Misalignment in policy updates across levels can lead to transient non-stationarity, which may affect stability in some scenarios. Furthermore, the use of a centralized high-level controller may limit decentralization in fully distributed deployments. Future work can explore

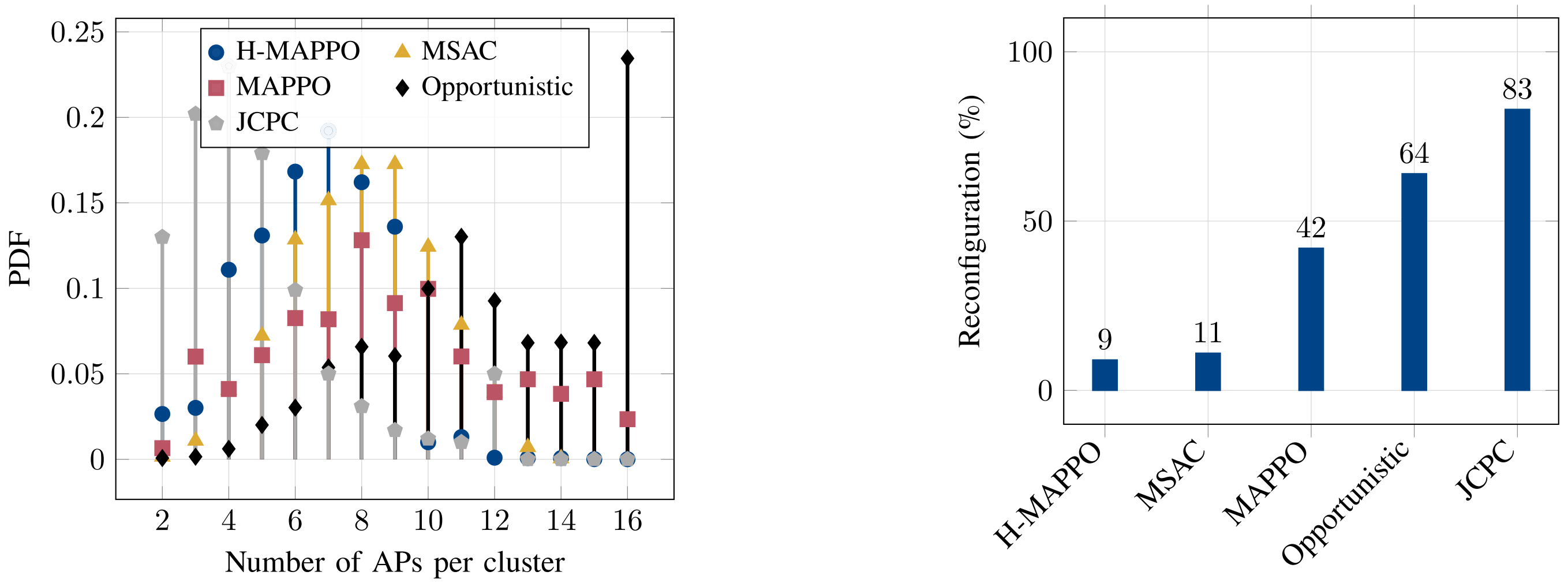

(a) Average cluster size for serving mobile UAVs.

(b) Cluster reconfiguration percentage of different methods.

Fig. 6. Numerical results showing the distribution of serving cluster sizes and the average percentage of cluster reconfigurations as users move.

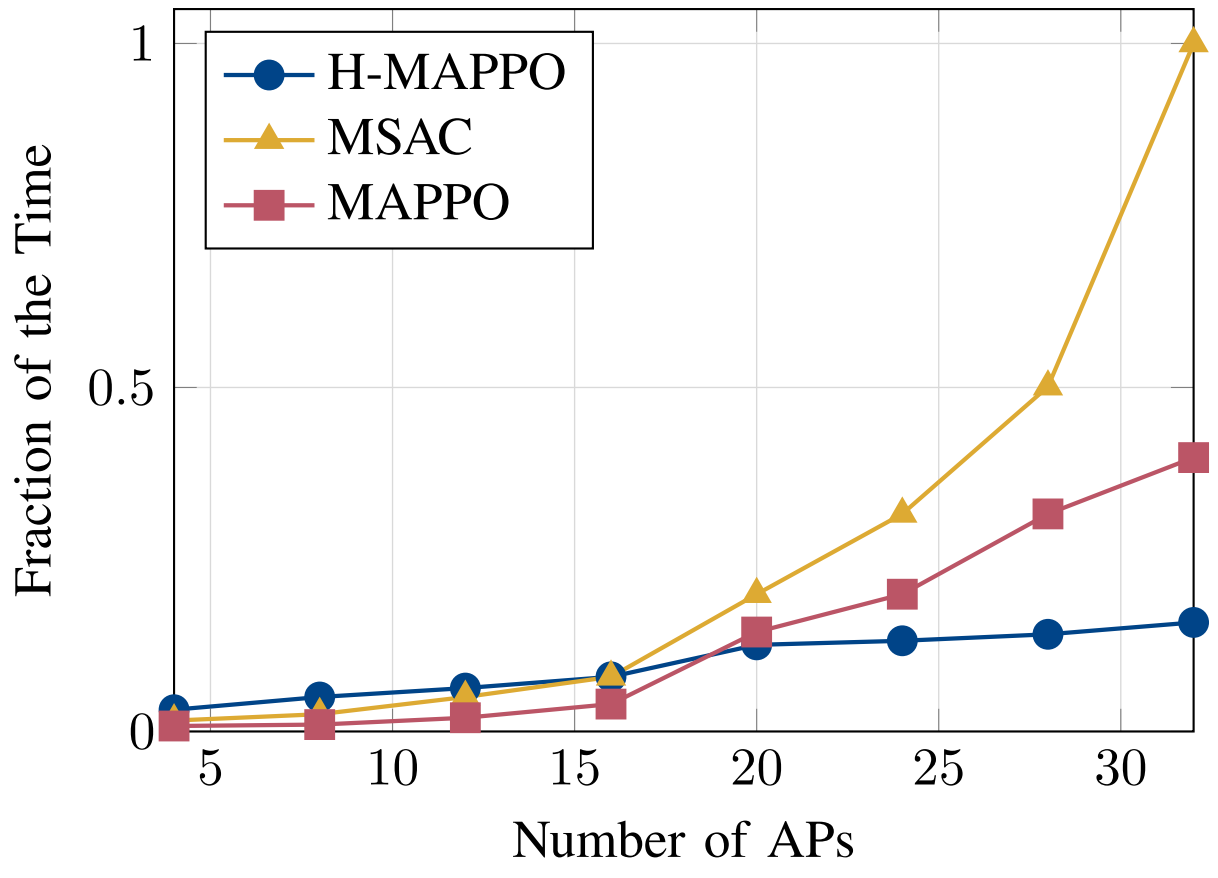

Fig. 7. Numerical results of the relative time needed to complete an episode during policy training.

communication-efficient variants or fully decentralized hierarchical formulations to mitigate these limitations.

## VI. Conclusion

In this paper, we have investigated the mobility management for multi-connectivity users in an wireless interference network under stringent reliability requirements. The mobility management problem involves joint dynamic cluster reconfiguration and energy-efficient power allocation with stringent QoS requirements. To solve it, we first divided the joint problem into two subproblems and propose a hierarchical two-layer H-MAPPO-based mobility management scheme. For the first layer of H-MAPPO, we have transformed the original multi-connectivity subproblem into a cluster reconfiguration updating problem, and leveraged the single agent PPO algorithm to output an optimized clustering action. For the second layer, each AP acts as an independent agent responsible for efficient power allocation to their assigned users using the MAPPO algorithm. The learning efficiency is improved through a novel action-observation transition mechanism that makes the convergence of the two layers possible.

Extensive simulation results verify the effectiveness of the proposed `H-MAPPO` scheme in reducing both outage probability and total transmit power. `H-MAPPO` achieves an outage probability near $10^{-6}$ and operates below $60\%$ of maximum transmit power in $90\%$ of instances, compared to only $30\%$ for opportunistic clustering methods. These results underscore its advantage in enhancing reliability and power efficiency for serving mobile UAVs.

In this work, we focused on user mobility within a single cloud's service area, with a single agent responsible for intra-cloud cluster reconfiguration. A natural extension of this work involves investigating scenarios where user mobility spans multiple clouds, necessitating inter-cloud cluster reconfiguration involving multiple agents.

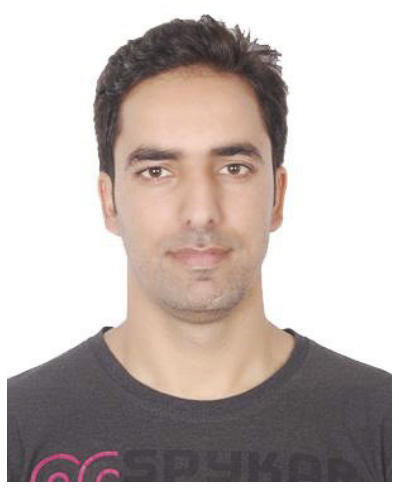

**Irshad A. Meer** (Member, IEEE) received the B.Tech. degree in electronics and communication engineering from the National Institute of Technology Srinagar, India, in 2014, the M.Sc. degree in telecommunication engineering from the Politecnico di Milano, Milan, Italy, in 2019, and the Ph.D. degree from the Department of Computer Science, KTH Royal Institute of Technology, Stockholm, Sweden, in 2025. He was a Visiting Student Research Collaborator with Princeton University, USA, from 2023 to 2024. He is currently a Post-Doctoral Fellow with the KTH Royal Institute of Technology. His research interests include machine learning for wireless communication systems, with a focus on explainable artificial intelligence for mobility management, UAV and drone-based communication networks, integrated terrestrial–non-terrestrial (TN-NTN) 6G architectures, resource allocation in cell-free networks, and AI-driven network automation.

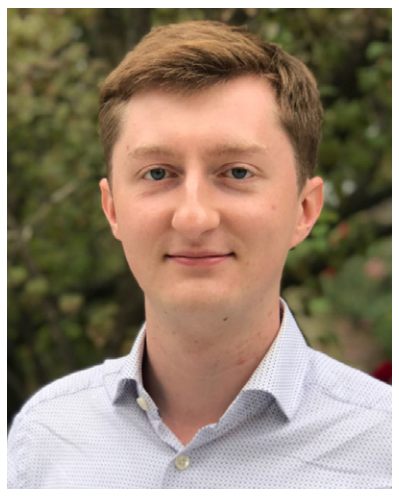

**Karl-Ludwig Besser** (Member, IEEE) received the Dipl.-Ing. degree in electrical engineering from Technische Universität Dresden, Germany, in 2018, and the Ph.D. degree in electrical engineering from Technische Universität Braunschweig, Germany, in 2022. In August 2018, he joined the Communications Theory Group with TU Dresden. From August 2019 to February 2023, he was with the Institute for Communications Technology, Technische Universität Braunschweig. From March 2023 to December 2024, he was as a Post-Doctoral Research Fellow with the Department of Electrical and Computer Engineering, Princeton University. Since January 2025, he has been an Assistant Professor with the Department of Electrical Engineering, Linköping University, Sweden. His research interests include ultra-reliable communications, resilience of communication systems, copulas, physical layer security, and the application of machine learning in communications.

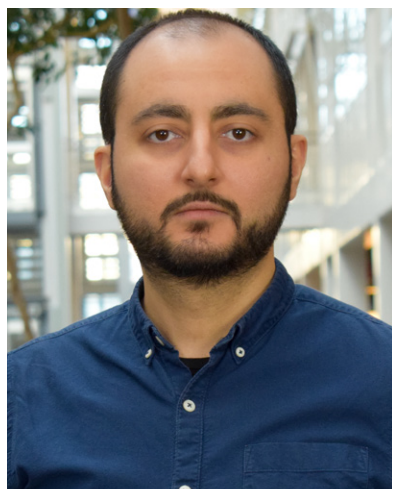

**Mustafa Ozger** (Member, IEEE) is currently an Assistant Professor with Aalborg University, Copenhagen, Denmark, and a Researcher with the KTH Royal Institute of Technology, Stockholm, Sweden. Previously, he was a Technical Coordinator of the Celtic-Next 6G-SKY Project. His research interests include aerial and vehicular networks, the Internet of Things, and AI-aided solutions for resource management problems in wireless networks and communications.

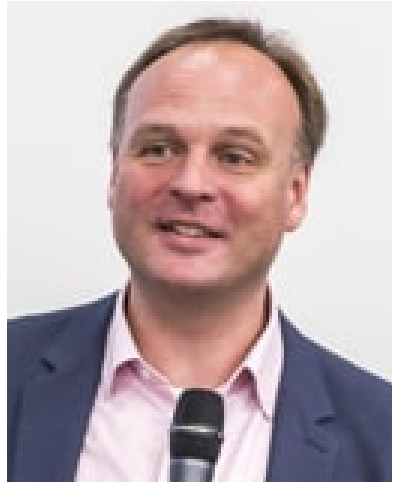

**Dominic A. Schupke** (Senior Member, IEEE) received the Dr.-Ing. degree (summa cum laude) in electrical engineering and information technology from RWTH Aachen University, Imperial College London, and the Technical University of Munich (TUM). He is currently a Research Leader in reliable communication networks, currently focusing on wireless communications at Airbus, Munich, Germany. He is also a Lecturer in network planning with TUM. Prior to Airbus, he was with Nokia, Siemens, and TUM. He is the author or co-author of more than 150 journals and conference papers (Google Scholar H-index 32). His recent research interests include aerospace networks.

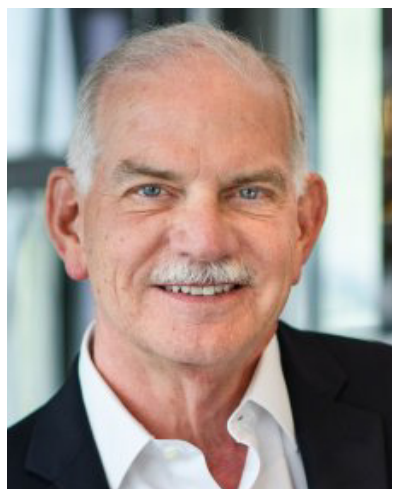

**H. Vincent Poor** (Life Fellow, IEEE) received the Ph.D. degree in EECS from Princeton University in 1977. From 1977 to 1990, he was with the Faculty member of the University of Illinois Urbana–Champaign. Since 1990, he has been with the Faculty Member with Princeton, where he is currently the Michael Henry Strater University Professor. From 2006 to 2016, he was the Dean of the Princeton's School of Engineering and Applied Science, and he has also held visiting appointments at several other universities, including most recently at Berkeley and Caltech. His research interests are in the areas of information theory, stochastic analysis and machine learning, and their applications in wireless networks, energy systems and related fields. Among his publications in these areas is the book *Machine Learning and Wireless Communications* (Cambridge University Press, 2022). He is a member of the National Academy of Engineering and the National Academy of Sciences and a Foreign Member of the Royal Society and other national and international academies. He received the IEEE Alexander Graham Bell Medal in 2017.

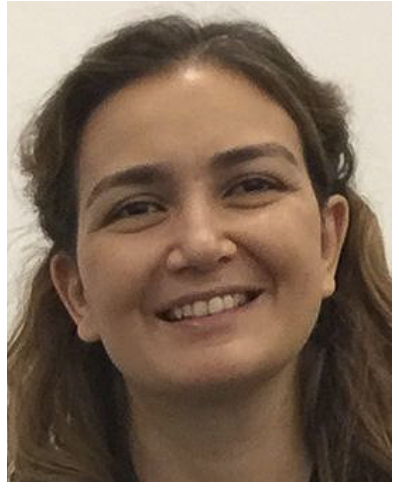

**Cicek Cavdar** (Member, IEEE) is currently a Professor with the KTH Royal Institute of Technology, Sweden, where she leads the Intelligent Network Systems Research Group and directs the SMART 6GSAT Satellite Communications Center. She is the Vice-Head of the Communication Systems Division and has played a leading role in numerous European Projects in aerospace communications and sustainable networking. Since 2013, she has been coordinated key EU initiatives on green communications, including 5GrEEn, SooGREEN, and the award-winning AI4Green. She also leads RAI-6Green and has driven industry-academic collaborations, such as ICARO-EU and 6G-SKY, where she has been serving as the Technical Coordinator. Her research focuses on the integration of non-terrestrial and terrestrial networks, cloud/edge computing, URLLC, and AI-driven sustainable network management.